\PassOptionsToPackage{unicode}{hyperref}
\PassOptionsToPackage{hyphens}{url}
\PassOptionsToPackage{dvipsnames,svgnames,x11names}{xcolor}
\documentclass[
]{article}
\usepackage{amsmath,amssymb}
\usepackage{lmodern}
\usepackage{iftex}
\ifPDFTeX
  \usepackage[T1]{fontenc}
  \usepackage[utf8]{inputenc}
  \usepackage{textcomp} % provide euro and other symbols
\else % if luatex or xetex
  \usepackage{unicode-math}
  \defaultfontfeatures{Scale=MatchLowercase}
  \defaultfontfeatures[\rmfamily]{Ligatures=TeX,Scale=1}
\fi
\IfFileExists{upquote.sty}{\usepackage{upquote}}{}
\IfFileExists{microtype.sty}{% use microtype if available
  \usepackage[]{microtype}
  \UseMicrotypeSet[protrusion]{basicmath} % disable protrusion for tt fonts
}{}
\makeatletter
\@ifundefined{KOMAClassName}{% if non-KOMA class
  \IfFileExists{parskip.sty}{%
    \usepackage{parskip}
  }{% else
    \setlength{\parindent}{0pt}
    \setlength{\parskip}{6pt plus 2pt minus 1pt}}
}{% if KOMA class
  \KOMAoptions{parskip=half}}
\makeatother
\usepackage{xcolor}
\NewDocumentCommand\citeproctext{}{}
\NewDocumentCommand\citeproc{mm}{%
  \begingroup\def\citeproctext{#2}\cite{#1}\endgroup}
\makeatletter
 \let\@cite@ofmt\@firstofone
 \def\@biblabel#1{}
 \def\@cite#1#2{{#1\if@tempswa , #2\fi}}
\makeatother
\newlength{\cslhangindent}
\newlength{\csllabelwidth}
\newenvironment{CSLReferences}[2] % #1 hanging-indent, #2 entry-spacing
 {\begin{list}{}{%
  \setlength{\itemindent}{0pt}
  \setlength{\leftmargin}{0pt}
  \setlength{\parsep}{0pt}
  \ifodd #1
   \setlength{\leftmargin}{\cslhangindent}
   \setlength{\itemindent}{-1\cslhangindent}
  \fi
  \setlength{\itemsep}{#2\baselineskip}}}
 {\end{list}}
\usepackage{calc}

\ifLuaTeX
\usepackage[bidi=basic]{babel}
\else
\usepackage[bidi=default]{babel}
\fi
\babelprovide[main,import]{american}
\def\languageshorthands#1{}
\ifLuaTeX
  \usepackage{selnolig}  % disable illegal ligatures
\fi
\IfFileExists{bookmark.sty}{\usepackage{bookmark}}{\usepackage{hyperref}}
\IfFileExists{xurl.sty}{\usepackage{xurl}}{} % add URL line breaks if available
\hypersetup{
  pdftitle={QuaRT: A toolkit for the exploration of quantum methods for
radiation transport},
  pdfauthor={Rasmit Devkota, John H. Wise},
  pdflang={en-US},
  colorlinks=true,
  linkcolor={Maroon},
  filecolor={Maroon},
  citecolor={Blue},
  urlcolor={Blue},
  pdfcreator={LaTeX via pandoc}}

\title{QuaRT: A toolkit for the exploration of quantum methods for
radiation transport}

\definecolor{c53baa1}{RGB}{83,186,161}
\definecolor{c202826}{RGB}{32,40,38}

\usepackage[affil-it]{authblk}
\usepackage{orcidlink}
\author[1,2,\ref{email}%
  ]{Rasmit Devkota%
    \,\orcidlink{0009-0009-3294-638X}\,%
    }
\author[1,3%
  ]{John H. Wise%
    \,\orcidlink{0000-0003-1173-8847}\,%
    }

\affil[1]{School of Physics, Georgia Institute of Technology, Atlanta,
GA%
  }
\affil[2]{School of Mathematics, Georgia Institute of Technology,
Atlanta, GA%
  }
\affil[3]{Center for Relativistic Astrophysics, Georgia Institute of
Technology, Atlanta, GA%
  }
\date{15 November 2025}

\renewcommand{\thefootnote}{\alph{footnote}}
\newcommand{\astfootnote}[1]{%
\let\oldthefootnote=\thefootnote%
\setcounter{footnote}{0}%
\renewcommand{\thefootnote}{\fnsymbol{footnote}}%
\footnote{#1}%
\let\thefootnote=\oldthefootnote%
}

\begin{document}
\maketitle

\astfootnote{\label{email}rdevkota3@gatech.edu}

\section{Summary}\label{summary}

\texttt{QuaRT} is a Python library for quantum simulation of radiative
transfer in astrophysical and cosmological problems.

The source code for \texttt{QuaRT} is available on
\href{https://github.com/RasmitDevkota/QuaRT/}{GitHub}. It can be
installed via \texttt{pip} from the
\href{https://pypi.org/project/PyQuaRT/}{\texttt{pypi} index}. Its
\href{https://pyquart.readthedocs.io/}{documentation} is hosted
publicly.

\section{Statement of need}\label{statement-of-need}

Computational cosmology, the use of simulations to study the evolution
of the universe, is a rapidly-growing field of research, driven largely
by the exponential increase in computing power following Moore's law.
Numerous codes (\citeproc{ref-BrummelSmith2019}{Brummel-Smith et al.,
2019}; \citeproc{ref-Bryan2014}{Bryan et al., 2014};
\citeproc{ref-Davis2012}{Davis et al., 2012};
\citeproc{ref-Hayes2003}{Hayes \& Norman, 2003};
\citeproc{ref-Iliev2006}{Ilian T. Iliev et al., 2006};
\citeproc{ref-Iliev2009}{Ilian T. Iliev et al., 2009};
\citeproc{ref-Jiang2014}{Jiang et al., 2014};
\citeproc{ref-Kannan2019}{Rahul Kannan et al., 2019};
\citeproc{ref-Kannan2021}{R. Kannan et al., 2021};
\citeproc{ref-OShea2015}{O'Shea et al., 2015}) have been written to
study questions about the early universe and to obtain a better
understanding of the plethora of observational results which have come
with new telescopes such as the James Webb Space Telescope
(\citeproc{ref-Adams2024}{Adams et al., 2024}). However, classical
high-performance computing hardware is slowly approaching the
fundamental quantum limit where electronics cannot be scaled down any
further (\citeproc{ref-Powell2008}{Powell, 2008}). Quantum computers
present a potential path for further scaling of physical simulations by
taking advantage of quantum phenomena such as superposition and
entanglement which enable new models of computation. Many quantum
algorithms have already been developed for the simulation of
cosmological problems (\citeproc{ref-Joseph2021}{Joseph et al., 2021},
\citeproc{ref-Joseph2022}{2022}; \citeproc{ref-Kaufman2019}{Kaufman et
al., 2019}; \citeproc{ref-Liu2021}{Liu \& Li, 2021};
\citeproc{ref-Mocz2021}{Mocz \& Szasz, 2021};
\citeproc{ref-Wang2024}{Wang \& Wu, 2024};
\citeproc{ref-Yamazaki2025}{Yamazaki et al., 2025}). Such simulations
must model physical processes such as radiation transport from stars,
magnetohydrodynamics of matter, gravitation between massive particles,
gas chemistry, and the formation of structures such as stars, black
holes, halos, and galaxies (e.g.,
\citeproc{ref-BrummelSmith2019}{Brummel-Smith et al., 2019};
\citeproc{ref-Hopkins2023}{Hopkins et al., 2023}). Of these, radiation
transport tends to be one of the most expensive steps due to the high
dimensionality of the problem, but it also the most difficult to develop
because of the lack of problems with analytical solutions
(\citeproc{ref-Iliev2006}{Ilian T. Iliev et al., 2006};
\citeproc{ref-Iliev2009}{Ilian T. Iliev et al., 2009}). Quantum
algorithms have been formulated for radiation transport, such as those
based on ray tracing (\citeproc{ref-Lu20221}{Lu \& Lin, 2022a},
\citeproc{ref-Lu20222}{2022b}, \citeproc{ref-Lu2023}{2023};
\citeproc{ref-Mosier2023}{Mosier, 2023};
\citeproc{ref-Santos2025}{Santos et al., 2025}), random walks
(\citeproc{ref-Lee2025}{Lee et al., 2025}), and other novel differential
equations solvers (\citeproc{ref-Gaitan2024}{Gaitan et al., 2024}).
Classical lattice Boltzmann methods (LBMs), which track the distribution
of a quantity on a grid with discretized propagation directions
(\citeproc{ref-McNamaraZanetti1988}{McNamara \& Zanetti, 1988}), have
already been applied extensively to study radiation transport
(\citeproc{ref-BindraPatil2012}{Bindra \& Patil, 2012};
\citeproc{ref-McCulloch2016}{McCulloch \& Bindra, 2016};
\citeproc{ref-Mink2020}{Mink et al., 2020};
\citeproc{ref-Olsen2025}{Olsen \& Rezzolla, 2025};
\citeproc{ref-Weih2020}{Weih et al., 2020}) and radiation hydrodynamics
(\citeproc{ref-Asahina2020}{Asahina et al., 2020}). Quantum LBMs have
also been constructed to study hydrodynamics
(\citeproc{ref-Budinski2021}{Budinski, 2021};
\citeproc{ref-Budinski2022}{Ljubomir, 2022};
\citeproc{ref-Wawrzyniak20252}{Wawrzyniak, Winter, Schmidt, Indiniger,
et al., 2025}; \citeproc{ref-Wawrzyniak20251}{Wawrzyniak, Winter,
Schmidt, Indinger, et al., 2025}) and radiation transport
(\citeproc{ref-Igarashi2024}{Igarashi et al., 2024}). These quantum LBMs
reduce the memory constraints of classical simulations by storing
information in quantum state amplitudes, the number of which grows
exponentially with the number of qubits, enabling the storage of data
with only logarithmic scaling with problem size. Individual simulation
steps can thus be made very high resolution and only the necessary
amount of data needs to be stored classically. However, existing quantum
LBMs are not suited for cosmological problems because such simulations
are typically non-scattering, but isotropic sources under stars are not
accurately resolved angularly by LBMs due to their discretized angular
structure. \texttt{QuaRT} features the first known implementation of a
quantum LBM which accurately resolves isotropic sources in
non-scattering media; it does so via a novel methodology which we refer
to as ``angular redistribution'', where radiation is redistributed
between angular directions based on the expected angular distribution.
This can even be done globally for an entire simulation domain with no
increase in computational complexity, enabling larger and more accurate
simulations of the evolution of the universe than currently possible.

\section{Functionality}\label{functionality}

\texttt{QuaRT} uses \texttt{Qiskit} for circuit construction and
\texttt{Qiskit\ Aer} for circuit execution
(\citeproc{ref-Javadi-Abhari2024}{Javadi-Abhari et al., 2024}).

The \texttt{qlbm\_rt} module features the \texttt{simulate} method which
is called to perform simulations with the lattice Boltzmann method. This
method constructs the full quantum circuit for each timestep of the
simulation and returns the lattice data.

The \texttt{qlbm\_circuits} module features constructors for the
necessary circuits for radiative transfer simulation in 1D, 2D, and 3D,
including a constructor for the novel angular redstribution step. These
constructors are called by the \texttt{simulate} method which composes
them to construct the full quantum circuit.

\texttt{QuaRT} features a variety of utility methods for both general
and quantum lattice Boltzmann methods in \texttt{lbm\_utils} and
\texttt{qlbm\_utils}, respectively. It also features analysis utilities
in the \texttt{analysis} module. These utilities are used by the
\texttt{simulate} method for problem setup and analysis.

The \texttt{test} module features a variety of common test cases used
for radiative transfer codes, including the isotropic source, opaque
cloud shadow, and crossing radiation beams tests. These tests
demonstrate the general correctness of the codebase, with a particular
emphasis on the performance of the angular redistribution methodology.

There is also a set of demo notebooks including a fully-classical
implementation for comparison with the quantum algorithm and some unit
tests of experimental features.

\section{Scholarly Work}\label{scholarly-work}

\texttt{QuaRT} is currently being used to study lattice Boltzmann
methods for radiative transfer (\citeproc{ref-Devkota2025}{Devkota \&
Wise, 2025}, in prep.).

\section{Acknowledgements}\label{acknowledgements}

JHW acknowledges support from NSF grants AST-2108020 and AST-2510197 and
NASA grant 80NSSC21K1053. This research was supported in part through
research cyberinfrastructure resources and services provided by the
Partnership for an Advanced Computing Environment (PACE) at the Georgia
Institute of Technology, Atlanta, Georgia, USA
(\citeproc{ref-PACE}{PACE, 2017}).

\section*{References}\label{references}
\addcontentsline{toc}{section}{References}

\phantomsection\label{refs}


\begin{CSLReferences}{1}{0}
\bibitem[\citeproctext]{ref-Adams2024}
Adams, N. J., Conselice, C. J., Austin, D., Harvey, T., Ferreira, L.,
Trussler, J., Juodžbalis, I., Li, Q., Windhorst, R., Cohen, S. H.,
Jansen, R. A., Summers, J., Tompkins, S., Driver, S. P., Robotham, A.,
D'Silva, J. C. J., Yan, H., Coe, D., Frye, B., \ldots{} Zitrin, A.
(2024). EPOCHS. II. The ultraviolet luminosity function from 7.5 \&lt; z
\&lt; 13.5 using 180 arcmin2 of deep, blank fields from the PEARLS
survey and public JWST data. \emph{The Astrophysical Journal},
\emph{965}(2), 169. \url{https://doi.org/10.3847/1538-4357/ad2a7b}

\bibitem[\citeproctext]{ref-Asahina2020}
Asahina, Y., Takahashi, H. R., \& Ohsuga, K. (2020). A numerical scheme
for general relativistic radiation magnetohydrodynamics based on solving
a grid-based boltzmann equation. \emph{The Astrophysical Journal},
\emph{901}(2), 96. \url{https://doi.org/10.3847/1538-4357/abaf51}

\bibitem[\citeproctext]{ref-BindraPatil2012}
Bindra, H., \& Patil, D. V. (2012). Radiative or neutron transport
modeling using a lattice boltzmann equation framework. \emph{Physical
Review E}, \emph{86}(1).
\url{https://doi.org/10.1103/physreve.86.016706}

\bibitem[\citeproctext]{ref-BrummelSmith2019}
Brummel-Smith, C., Bryan, G., Butsky, I., Corlies, L., Emerick, A.,
Forbes, J., Fujimoto, Y., Goldbaum, N., Grete, P., Hummels, C., Kim, J.,
Koh, D., Li, M., Li, Y., Li, X., O'Shea, B., Peeples, M., Regan, J.,
Salem, M., \ldots{} Zhao, F. (2019). ENZO: An adaptive mesh refinement
code for astrophysics (version 2.6). \emph{Journal of Open Source
Software}, \emph{4}(42), 1636. \url{https://doi.org/10.21105/joss.01636}

\bibitem[\citeproctext]{ref-Bryan2014}
Bryan, G. L., Norman, M. L., O'Shea, B. W., Abel, T., Wise, J. H., Turk,
M. J., Reynolds, D. R., Collins, D. C., Wang, P., Skillman, S. W.,
Smith, B., Harkness, R. P., Bordner, J., Kim, J., Kuhlen, M., Xu, H.,
Goldbaum, N., Hummels, C., Kritsuk, A. G., \ldots{} Li, Y. (2014). ENZO:
AN ADAPTIVE MESH REFINEMENT CODE FOR ASTROPHYSICS. \emph{The
Astrophysical Journal Supplement Series}, \emph{211}(2), 19.
\url{https://doi.org/10.1088/0067-0049/211/2/19}

\bibitem[\citeproctext]{ref-Budinski2021}
Budinski, L. (2021). Quantum algorithm for the advection--diffusion
equation simulated with the lattice boltzmann method. \emph{Quantum
Information Processing}, \emph{20}(2).
\url{https://doi.org/10.1007/s11128-021-02996-3}

\bibitem[\citeproctext]{ref-Davis2012}
Davis, S. W., Stone, J. M., \& Jiang, Y.-F. (2012). A RADIATION TRANSFER
SOLVER FOR ATHENA USING SHORT CHARACTERISTICS. \emph{The Astrophysical
Journal Supplement Series}, \emph{199}(1), 9.
\url{https://doi.org/10.1088/0067-0049/199/1/9}

\bibitem[\citeproctext]{ref-Devkota2025}
Devkota, R., \& Wise, J. H. (2025). {A novel quantum lattice Boltzmann
methodology for cosmological radiation transport}. \emph{Phys. Rev. D}.

\bibitem[\citeproctext]{ref-Gaitan2024}
Gaitan, F., Graziani, F., \& Porter, M. D. (2024). Simulating nonlinear
radiation diffusion through quantum computing. \emph{International
Journal of Theoretical Physics}, \emph{63}(10).
\url{https://doi.org/10.1007/s10773-024-05800-x}

\bibitem[\citeproctext]{ref-Hayes2003}
Hayes, J. C., \& Norman, M. L. (2003). Beyond flux‐limited diffusion:
Parallel algorithms for multidimensional radiation hydrodynamics.
\emph{The Astrophysical Journal Supplement Series}, \emph{147}(1),
197--220. \url{https://doi.org/10.1086/374658}

\bibitem[\citeproctext]{ref-Hopkins2023}
Hopkins, P. F., Wetzel, A., Wheeler, C., Sanderson, R., Grudić, M. Y.,
Sameie, O., Boylan-Kolchin, M., Orr, M., Ma, X., Faucher-Giguère, C.-A.,
Kereš, D., Quataert, E., Su, K.-Y., Moreno, J., Feldmann, R., Bullock,
J. S., Loebman, S. R., Anglés-Alcázar, D., Stern, J., \ldots{} Hayward,
C. C. (2023). {FIRE-3: updated stellar evolution models, yields, and
microphysics and fitting functions for applications in galaxy
simulations}. \emph{519}(2), 3154--3181.
\url{https://doi.org/10.1093/mnras/stac3489}

\bibitem[\citeproctext]{ref-Igarashi2024}
Igarashi, A., Kadowaki, T., \& Kawabata, S. (2024). Quantum algorithm
for the radiative-transfer equation. \emph{Physical Review Applied},
\emph{21}(3).

\bibitem[\citeproctext]{ref-Iliev2006}
Iliev, Ilian T., Ciardi, B., Alvarez, M. A., Maselli, A., Ferrara, A.,
Gnedin, N. Y., Mellema, G., Nakamoto, T., Norman, M. L., Razoumov, A.
O., Rijkhorst, E.-J., Ritzerveld, J., Shapiro, P. R., Susa, H., Umemura,
M., \& Whalen, D. J. (2006). Cosmological radiative transfer codes
comparison project -- i. The static density field tests. \emph{Monthly
Notices of the Royal Astronomical Society}, \emph{371}(3), 1057--1086.
\url{https://doi.org/10.1111/j.1365-2966.2006.10775.x}

\bibitem[\citeproctext]{ref-Iliev2009}
Iliev, Ilian T., Whalen, D., Mellema, G., Ahn, K., Baek, S., Gnedin, N.
Y., Kravtsov, A. V., Norman, M., Raicevic, M., Reynolds, D. R., Sato,
D., Shapiro, P. R., Semelin, B., Smidt, J., Susa, H., Theuns, T., \&
Umemura, M. (2009). Cosmological radiative transfer comparison project
II. The radiation-hydrodynamic tests. \emph{Mon. Not. R. Astron. Soc.},
\emph{400}(3).

\bibitem[\citeproctext]{ref-Javadi-Abhari2024}
Javadi-Abhari, A., Treinish, M., Krsulich, K., Wood, C. J., Lishman, J.,
Gacon, J., Martiel, S., Nation, P. D., Bishop, L. S., Cross, A. W.,
Johnson, B. R., \& Gambetta, J. M. (2024). \emph{Quantum computing with
{Q}iskit}. \url{https://doi.org/10.48550/arXiv.2405.08810}

\bibitem[\citeproctext]{ref-Jiang2014}
Jiang, Y.-F., Stone, J. M., \& Davis, S. W. (2014). AN ALGORITHM FOR
RADIATION MAGNETOHYDRODYNAMICS BASED ON SOLVING THE TIME-DEPENDENT
TRANSFER EQUATION. \emph{The Astrophysical Journal Supplement Series},
\emph{213}(1), 7. \url{https://doi.org/10.1088/0067-0049/213/1/7}

\bibitem[\citeproctext]{ref-Joseph2021}
Joseph, A., Varela, J.-P., Watts, M. P., White, T., Feng, Y., Hassan,
M., \& McGuigan, M. (2021). \emph{Quantum computing for inflationary,
dark energy and dark matter cosmology}. arXiv.
\url{https://doi.org/10.48550/ARXIV.2105.13849}

\bibitem[\citeproctext]{ref-Joseph2022}
Joseph, A., White, T., Chandra, V., \& McGuigan, M. (2022).
\emph{Quantum computing of schwarzschild-de sitter black holes and
kantowski-sachs cosmology}. arXiv.
\url{https://doi.org/10.48550/ARXIV.2202.09906}

\bibitem[\citeproctext]{ref-Kannan2021}
Kannan, R., Garaldi, E., Smith, A., Pakmor, R., Springel, V.,
Vogelsberger, M., \& Hernquist, L. (2021). Introducing the
\textless scp\textgreater thesan\textless/scp\textgreater{} project:
Radiation-magnetohydrodynamic simulations of the epoch of reionization.
\emph{Monthly Notices of the Royal Astronomical Society}, \emph{511}(3),
4005--4030. \url{https://doi.org/10.1093/mnras/stab3710}

\bibitem[\citeproctext]{ref-Kannan2019}
Kannan, Rahul, Vogelsberger, M., Marinacci, F., McKinnon, R., Pakmor,
R., \& Springel, V. (2019).
\textless Scp\textgreater arepo-rt\textless/scp\textgreater: Radiation
hydrodynamics on a moving mesh. \emph{Monthly Notices of the Royal
Astronomical Society}, \emph{485}(1), 117--149.
\url{https://doi.org/10.1093/mnras/stz287}

\bibitem[\citeproctext]{ref-Kaufman2019}
Kaufman, A., Sundy, D., \& McGuigan, M. (2019). Quantum computation for
early universe cosmology. \emph{2019 New York Scientific Data Summit
(NYSDS)}, 1--6. \url{https://doi.org/10.1109/nysds.2019.8909801}

\bibitem[\citeproctext]{ref-Lee2025}
Lee, E., Lee, S., \& Kim, S. (2025). Quantum walk based monte carlo
simulation for photon interaction cross sections. \emph{Physical Review
D}, \emph{111}(11). \url{https://doi.org/10.1103/physrevd.111.116001}

\bibitem[\citeproctext]{ref-Liu2021}
Liu, J., \& Li, Y.-Z. (2021). Quantum simulation of cosmic inflation.
\emph{Physical Review D}, \emph{104}(8).
\url{https://doi.org/10.1103/physrevd.104.086013}

\bibitem[\citeproctext]{ref-Budinski2022}
Ljubomir, B. (2022). Quantum algorithm for the navier--stokes equations
by using the streamfunction-vorticity formulation and the lattice
boltzmann method. \emph{International Journal of Quantum Information},
\emph{20}(02). \url{https://doi.org/10.1142/s0219749921500398}

\bibitem[\citeproctext]{ref-Lu20221}
Lu, X., \& Lin, H. (2022a). \emph{A framework for quantum ray tracing}.
\url{https://doi.org/10.48550/ARXIV.2203.15451}

\bibitem[\citeproctext]{ref-Lu20222}
Lu, X., \& Lin, H. (2022b). Quantum ray tracing with simulation.
\emph{SPIN}, \emph{12}(04).
\url{https://doi.org/10.1142/s2010324722500308}

\bibitem[\citeproctext]{ref-Lu2023}
Lu, X., \& Lin, H. (2023). Improved quantum supersampling for quantum
ray tracing. \emph{Quantum Information Processing}, \emph{22}(10).
\url{https://doi.org/10.1007/s11128-023-04114-x}

\bibitem[\citeproctext]{ref-McCulloch2016}
McCulloch, R., \& Bindra, H. (2016). Coupled radiative and conjugate
heat transfer in participating media using lattice boltzmann methods.
\emph{Computers \&Amp; Fluids}, \emph{124}, 261--269.
\url{https://doi.org/10.1016/j.compfluid.2015.05.024}

\bibitem[\citeproctext]{ref-McNamaraZanetti1988}
McNamara, G. R., \& Zanetti, G. (1988). Use of the boltzmann equation to
simulate lattice-gas automata. \emph{Phys. Rev. Lett.}, \emph{61},
2332--2335. \url{https://doi.org/10.1103/PhysRevLett.61.2332}

\bibitem[\citeproctext]{ref-Mink2020}
Mink, A., McHardy, C., Bressel, L., Rauh, C., \& Krause, M. J. (2020).
Radiative transfer lattice boltzmann methods: 3D models and their
performance in different regimes of radiative transfer. \emph{Journal of
Quantitative Spectroscopy and Radiative Transfer}, \emph{243}, 106810.
\url{https://doi.org/10.1016/j.jqsrt.2019.106810}

\bibitem[\citeproctext]{ref-Mocz2021}
Mocz, P., \& Szasz, A. (2021). Toward cosmological simulations of dark
matter on quantum computers. \emph{The Astrophysical Journal},
\emph{910}(1), 29. \url{https://doi.org/10.3847/1538-4357/abe6ac}

\bibitem[\citeproctext]{ref-Mosier2023}
Mosier, L. (2023). \emph{Quantum ray marching: Reformulating light
transport for quantum computers} {[}Master\textquotesingle s Thesis{]}.
University of Waterloo.

\bibitem[\citeproctext]{ref-OShea2015}
O'Shea, B. W., Wise, J. H., Xu, H., \& Norman, M. L. (2015). PROBING THE
ULTRAVIOLET LUMINOSITY FUNCTION OF THE EARLIEST GALAXIES WITH THE
RENAISSANCE SIMULATIONS. \emph{The Astrophysical Journal},
\emph{807}(1), L12. \url{https://doi.org/10.1088/2041-8205/807/1/l12}

\bibitem[\citeproctext]{ref-Olsen2025}
Olsen, T., \& Rezzolla, L. (2025). General-relativistic
lattice-boltzmann method for radiation transport. \emph{Monthly Notices
of the Royal Astronomical Society}, \emph{541}(2), 1305--1328.
\url{https://doi.org/10.1093/mnras/staf992}

\bibitem[\citeproctext]{ref-PACE}
PACE. (2017). \emph{{P}artnership for an {A}dvanced {C}omputing
{E}nvironment ({PACE})}. \url{http://www.pace.gatech.edu}

\bibitem[\citeproctext]{ref-Powell2008}
Powell, J. R. (2008). The quantum limit to moore's law.
\emph{Proceedings of the IEEE}, \emph{96}(8), 1247--1248.
\url{https://doi.org/10.1109/JPROC.2008.925411}

\bibitem[\citeproctext]{ref-Santos2025}
Santos, L. P., Bashford-Rogers, T., Barbosa, J., \& Navrátil, P. (2025).
Towards quantum ray tracing. \emph{IEEE Transactions on Visualization
and Computer Graphics}, \emph{31}(4), 2223--2234.
\url{https://doi.org/10.1109/tvcg.2024.3386103}

\bibitem[\citeproctext]{ref-Wang2024}
Wang, C.-C. E., \& Wu, J.-H. P. (2024). \emph{Quantum cosmology on
quantum computer}. \url{https://doi.org/10.48550/ARXIV.2410.22485}

\bibitem[\citeproctext]{ref-Wawrzyniak20251}
Wawrzyniak, D., Winter, J., Schmidt, S., Indinger, T., Janßen, C. F.,
Schramm, U., \& Adams, N. A. (2025). A quantum algorithm for the
lattice-boltzmann method advection-diffusion equation. \emph{Computer
Physics Communications}, \emph{306}, 109373.
\url{https://doi.org/10.1016/j.cpc.2024.109373}

\bibitem[\citeproctext]{ref-Wawrzyniak20252}
Wawrzyniak, D., Winter, J., Schmidt, S., Indiniger, T., Janßen, C. F.,
Schramm, U., \& Adams, N. A. (2025). \emph{Dynamic circuits for the
quantum lattice-boltzmann method}.
\url{https://doi.org/10.48550/ARXIV.2502.02131}

\bibitem[\citeproctext]{ref-Weih2020}
Weih, L. R., Gabbana, A., Simeoni, D., Rezzolla, L., Succi, S., \&
Tripiccione, R. (2020). Beyond moments: Relativistic lattice boltzmann
methods for radiative transport in computational astrophysics.
\emph{Monthly Notices of the Royal Astronomical Society}, \emph{498}(3),
3374--3394. \url{https://doi.org/10.1093/mnras/staa2575}

\bibitem[\citeproctext]{ref-Yamazaki2025}
Yamazaki, S., Uchida, F., Fujisawa, K., Miyamoto, K., \& Yoshida, N.
(2025). Quantum algorithm for collisionless boltzmann simulation of
self-gravitating systems. \emph{Computers \&Amp; Fluids}, \emph{288},
106527. \url{https://doi.org/10.1016/j.compfluid.2024.106527}

\end{CSLReferences}
\end{document}